\documentclass[11pt,oneside]{article}

\usepackage{geometry} % see geometry.pdf on how to lay out the page. There's lots.
\usepackage{eurosym}

\usepackage{times}

\usepackage{natbib}
\usepackage{amsfonts}
\usepackage{amsthm}

\usepackage{graphicx}

\usepackage{epsfig}
\usepackage{subfigure}
\usepackage{amsmath}    % need for subequations
\usepackage{longtable}
\usepackage{dcolumn}
\usepackage{amssymb}

\usepackage{wrapfig}

\DeclareGraphicsExtensions{.jpg,.eps,.EPS,.png,.pdf,.ps}

\usepackage{colordvi}
\usepackage[usenames,dvipsnames]{xcolor}

\RequirePackage[colorlinks=true
,urlcolor=blue
,anchorcolor=blue
,citecolor=blue
,filecolor=blue
,linkcolor=blue
,menucolor=blue
,pagecolor=blue
,linktocpage=true
,pdfproducer=medialab
]{hyperref}

\title{Approximate analytical solutions to the condensation-coagulation equation of aerosols}

\author{Naftali Smith$^{1}$, Nir J. Shaviv$^{1,2}$, Henrik Svensmark$^{3}$ \\
\normalsize $^{1}$Racah Institute of Physics, Hebrew University of Jerusalem, Israel \\
\normalsize $^{2}$The Institute for Advanced study, Princeton NJ 08540, USA\\
\normalsize $^{3}$ National Space Institute, Technical University of Denmark, \\ Elektrovej, Bygn. 328, 2800 Lyngby, Denmark
}

\date{}

\begin{document}

\maketitle

\parskip 5pt 

\begin{abstract}
We present analytical solutions to the steady state injection-condensation-coagulation equation of aerosols in the atmosphere. These solutions are appropriate under different limits but more general than previously derived analytical solutions. For example, we provide an analytic solution to the coagulation limit plus a condensation correction. Our solutions are then compared with numerical results. We show that the solutions can be used to estimate the sensitivity of the cloud condensation nuclei number density to the nucleation rate of small condensation nuclei and to changes in the formation rate of sulfuric acid.  
\end{abstract}

\vfill \eject

\section{Introduction}
\label{sec:intro}

Aerosols in Earth's atmosphere are important for the climate systems as they are required to form clouds. As a consequence, different aerosol characteristics translate into different cloud properties, such as different radiative forcing, different life time and precipitation. Since a large part of the characteristics of aerosols can be described by their size distribution, solving for the distribution is of particular interest.

For a spatially homogeneous aerosol distribution, the size distribution can be described with a density function $n(v,t)$ where $v$ is the aerosol size, such that $\int_{v_1}^{v_2} n(v,t) dv$ is the total number of aerosols {\em per unit volume}, with volumes between $v_1$ and $v_2$.

The main equation describing the temporal evolution of $n(v,t)$ is the coagulation-condensation equation \citep[e.g.,][]{Peterson1978,Seinfeld2006}:
\begin{eqnarray}
 \frac{\partial n}{\partial t}+\frac{\partial}{\partial v}\left[I\left(v,t\right)n\right] &=&  
\frac{1}{2}\int_{0}^{v}\beta\left(v-\tilde{v},\tilde{v}\right)n\left(v-\tilde{v},t\right)n\left(\tilde{v},t\right)d\tilde{v} \nonumber \\
& & -\int_{0}^{\infty}\beta\left(v,\tilde{v}\right)n\left(v,t\right)n\left(\tilde{v},t\right)d\tilde{v}-R(v,t)+S(v,t),
\label{eq:FullCCE}
\end{eqnarray}
where $ I ( v , t ) = dv/dt$ is the growth rate of
a particle of volume $v$, $\beta (v_1,v_2)$ is the coagulation
coefficient for particles with volumes $v_1$
and $v_2$, $R(v,t)$ is the rate with which particles are removed from the system, and $S(v,t)$ is the
nucleation rate of fresh particles. It is of course similar to the \cite{Smol} equation, with the addition of the condensation term. 

\cite{Klett1975} provided an analytical solution to the above equation, but without the condensation term, for a few specific cases by means of a Laplace transform. For example, he solved the aerosol distribution for the case of a mass independent coagulation coefficient. 

Later, \cite{Ram1976} provided an analytic solution to the more general problem which includes condensation. Like Klett, they have shown that power laws can provide approximate solutions in different regimes. 
\cite{Peterson1978} elaborated on the above and provided more general solutions. For example, they solved the time dependent problem. However, they confined themselves to cases in which the condensation is either constant or linearly dependent on the aerosol volume.  

It should be noted that an equation  similar to the aerosol condensation/coagulation equation appears in other contexts  as well. First, the original  coagulation equation (with no condensation) was formulated by \cite{Smol} to describe a colloidal fluid. A second example is that of cloud raindrops which behave similarly to the smaller sized aerosols with the main exception that their coagulation equation includes raindrop fragmentation \citep[e.g., see review by][]{Beheng}, while outside the terrestrial settings, 
\cite{Birnstiel2011} solve a coagulation/fragmentation equation describing dust grains in the interstellar medium.

We begin in \S\ref{sec:problem} by writing the coagulation-condensation equation in a dimensionless form and the approximations we assume in the present analysis. This will define the problem that we solve.
In  \S\ref{sec:CoalescenceLimit},  we first solve the problem while discarding the condensation term, and by doing so we arrive at the same solution found in \cite{Klett1975} by different means. This solution is expanded by treating the condensation term as a perturbation. The new analytic solution is appropriate for large aerosols where the dominant process is coagulation.
In \S\ref{sec:CondensationLimit},  we solve the opposite limit, first, when the coagulation term is altogether negligible, and then, when its effects are approximated. This solution is appropriate for small aerosols where the dominant process is condensation.
In \S\ref{sec:Numerical} we compare our results to a numerical calculation (the details of which can be found in the appendix).
We end in \S\ref{sec:Discussion} by discussing the significance of our theoretical results.

\section{The problem}
\label{sec:problem}

\def\Kn{\mathrm{Kn}}
%\subsection{original form}

Although the general problem of aerosol growth, eq.\ \ref{eq:FullCCE}, is time dependent. We will concentrate in the present work on the time-independent case; we will assume a steady-state solution $n(v)$.

As in \cite{Peterson1978}, we will assume a constant coagulation coefficient $\beta = \beta_0$ and $I\left(v,t\right)=\sigma v^{\gamma}$ where $0 \le \gamma \le 1$ , and we will emphasize the case $\gamma={1}/{3}$ (which is appropriate for the ``continuum regime") and $\gamma={2}/{3}$ (which is appropriate for the ``kinetic regime"). The continuum and kinetic regimes are defined by the Knudsen number $\Kn=\lambda/R$, where $\lambda$ is the mean free path, and $R$ is the radius of the paricle. The continuum regime corresponds to $\Kn\ll 1$, and the kinetic regime corresponds to $\Kn\gg 1$. The transition between these two regimes actually takes place for  $\Kn\approx 0.1$ (at which point the interaction rate is half the kinetic regime), occurs when $R \approx \lambda \approx 0.5\mu$m (for the earth's atmosphere at sea level, e.g., see \citealt{Seinfeld2006})
 
We also assume that the timescale for removal of particles from the system is much larger than the time necessary to arrive at the steady-state solution. Therefore we will discard the removal term altogether.

Finally, we assume a delta function source $S\left(v,t\right)=S_{0}\delta\left(v-v_{1}\right)$, again following \cite{Peterson1978}. As we shall see below, we will also assume that $v_1$ is much smaller than some characteristic volume that will be defined later.

We therefore obtain the following integro-differential equation for the steady-state distribution:
\begin{equation}
\label{eq:DimensionEquation}
\frac{d}{dv}\left[\sigma v^{\gamma}n\left(v\right)\right]=\frac{1}{2}\int_{0}^{v}\beta_{0}n\left(\tilde{v}\right)n\left(v-\tilde{v}\right)d\tilde{v}-\int_{0}^{\infty}\beta_{0}n\left(\tilde{v}\right)n\left(v\right)d\tilde{v}+S_{0}\delta\left(v-v_{1}\right).
\end{equation}

\subsection{General Kernel}
\label{sec:generalkernel}

Although we solve in the present work the coagulation equation while assuming a constant coagulation coefficient, we note that the any solution under this assumption can be immediately generalized to coagulations having kernels of the form $\beta\left(v,\tilde{v}\right)=\beta_{1}v^{\alpha}\tilde{v}^{\alpha}$.  

Here, the coagulation equation has the form
\begin{equation}
\frac{d}{dv}\left[\sigma v^{\gamma}n\left(v\right)\right]=\frac{1}{2}\int_{0}^{v}\beta_{1}\tilde{v}^{\alpha}\left(v-\tilde{v}\right)^{\alpha}n(\tilde{v})n\left(v-\tilde{v}\right)d\tilde{v}-\int_{0}^{\infty}\beta_{1}\tilde{v}^{\alpha}v^{\alpha}n(\tilde{v})n\left(v\right)d\tilde{v}+S_{0}\delta\left(v-v_{1}\right).
\end{equation}
 
To reduce it to the fixed kernel form, we define
\begin{equation}
f\left(v\right) \equiv v^{\alpha}n\left(v\right),
 \end{equation}
giving rise to the following equation for $f$:
\begin{equation}
\frac{d}{dv}\left[\sigma v^{\gamma-\alpha}f\left(v\right)\right]=\frac{1}{2}\int_{0}^{v}\beta_{1}f(\tilde{v})f\left(v-\tilde{v}\right)d\tilde{v}-\int_{0}^{\infty}\beta_{1}f(\tilde{v})f\left(v\right)d\tilde{v}+S_{0}\delta\left(v-v_{1}\right).
 \end{equation}
This equation is the same as eq.\ \ref{eq:DimensionEquation}, but with a different value of $\gamma$, that is $\gamma \rightarrow \gamma-\alpha$.

\subsection{Dimensionless Form}
\label{sec:dimensionless}

Before we solve eq.\ \ref{eq:DimensionEquation}, we recast it in a dimensionless form. We begin by defining 
 $\chi_{0}$ as the total number of particles in the system, that is, 
$ \chi_{0}\equiv \int_{0}^{\infty}n\left(v\right)dv
$. The differential equation which describes its temporal dependence 
is \citep{Klett1975}:
\begin{equation}
\frac{d\chi_{0}}{dt}=-\frac{1}{2}\beta_{0}\chi_{0}^{2}+S_{0},
\end{equation}
which gives a steady state solution of 
$ \chi_{0}=\sqrt{{2S_{0}}/{\beta_{0}}}$.

Next, we define the characteristic volume as 
\begin{equation}
v_{2} \equiv \left(\frac{\sigma^{2}}{S_{0}\beta_{0}}\right)^{{1}/{\left(2-2\gamma\right)}}
 \end{equation}
and characteristic particle density to be 
\begin{equation}
n_{0} \equiv \frac{\sqrt{{2S_{0}}/{\beta_{0}}}}{v_{2}}.
 \end{equation}
We also defined a dimensionless time as $\tau \equiv {t}/{\sqrt{{2}/{S_{0}\beta_{0}}}}$.

Using these characteristics quantities, we can define the dimensionless variables as
$x \equiv {v}/{v_{2}}$, $x_{1}\equiv {v_{1}}/{v_{2}}$, and $y\equiv {n}/{n_{0}}$, such that the time dependent version of eq.\ \ref{eq:DimensionEquation} becomes
\begin{equation}
\label{eq:DimensionlessEquation}
\frac{\partial y}{\partial\tau}+\sqrt{2}\frac{d}{dx}\left[x^{\gamma}y\left(x\right)\right]=\int_{0}^{x}y\left(\tilde{x}\right)y\left(x-\tilde{x}\right)d\tilde{x}-2y\left(x\right)+\delta\left(x-x_{1}\right).
 \end{equation} 

Note the fact that in steady state:
\begin{equation}
\int_{0}^{\infty}y\left(x\right)dx=1.
 \end{equation} 

So the steady state equation can also be written in the form
\begin{equation}
\sqrt{2}\frac{d}{dx}\left[x^{\gamma}y\left(x\right)\right]=\int_{0}^{x}y\left(\tilde{x}\right)y\left(x-\tilde{x}\right)d\tilde{x}-2y\left(x\right)\int_{0}^{\infty}y\left(\tilde{x}\right)d\tilde{x}+\delta\left(x-x_{1}\right)
 \end{equation} 
 which is closer to the form of equation\ \ref{eq:DimensionEquation}. We will assume throughout this paper that $v_1 \ll v_2$ , which translates to $x_1 \ll 1$ in the dimensionless quantities.

The amount of condensable matter is affected by the processes of nucleation and condensation. Most of this matter is sulphuric acid (SA) and water which condenses with it (though over the oceans methanesulfonic acid, MSA, could also be important). The equation which describes the change of the condensable material is:
\begin{equation}
\frac{dM_{SA}}{dt}=\epsilon_{m}-\int_{v_{1}}^{\infty}\sigma v^{\gamma}\frac{dn}{dv}dv-S_{0}v_{1}.
 \end{equation}
where $\epsilon_m$
is the rate of creation of sulphuric acid (or MSA) by an external source, together with the water which lcondenses with it.  Therefore, in steady-state we have:
\begin{equation}
 \epsilon_{m}=\int_{v_{1}}^{\infty}\sigma v^{\gamma}\frac{dn}{dv}dv-S_{0}v_{1}.
 \end{equation}

Later on we will discard the $S_{0}v_{1}$
term as it is smaller than the first term (under the assumption $v_{1} \ll v_{2}$).

\section{The coalescence limit}
\label{sec:CoalescenceLimit}

\subsection{Rough approximation}

A full solution to the coalescence equation will be presented in \S\ref{sec:fullSolution}. However, before we do so, it is worthwhile to derive the general behavior from very simple arguments. This will not provide a solution with accurate normalization constants, but it encapsulates the underlying physics and therefore it provides the correct power law.

We start with eq.\ \ref{eq:DimensionEquation}, and define $\epsilon_m\left(v\right)$ to be the rate of change in the total volume of particles of size $v$ or larger, that is,
\begin{equation}
\epsilon_m\left(v\right) \equiv \int_{v}^{\infty}\frac{\partial}{\partial t}n(\tilde{v})\tilde{v}d\tilde{v} .
\end{equation}
We shall try to calculate this quantity crudely, and then require it to be independent of $v$, as we expect it to be in steady state. We shall first assume no condensation, and later generalize to the case also having condensation as a perturbation. 

%%%%%%%%%%%%%%%%%%%%%%%%%%%%%%%%%%%%%%%%%%%%%%%%%%%%%%%%%%%%%%%%
\begin{figure}[h]
\centering
\includegraphics[width=0.35\columnwidth]{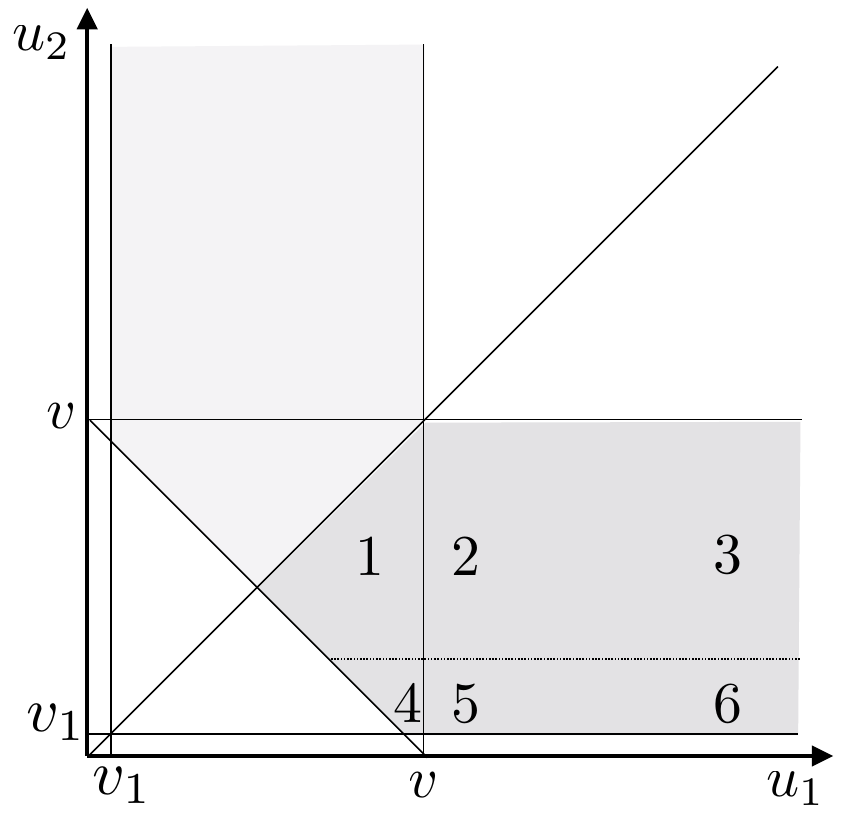}
\caption{$\epsilon_m$ is estimated by adding the contribution from the shaded area in the $(u_1,u_2)$ plane---the volumes of the coagulating aerosols. Unshaded regions don't contribute to $\epsilon_m$ either because they don't generate more volume above $v$, or because there are no particles in those regions (with $u_1$ or $u_2 < v$. In each region, the contribution to the integral behaves differently and therefore its typical value is different, as is summarized in table\ \ref{table:zones_u1_u2}. The light shaded area is a mirror of the darker shaded one. Instead of counting it separately, one can simply double the contribution from the darker region. }
 \label{fig:u1_u2_zones}
\end{figure}
%%%%%%%%%%%%%%%%%%%%%%%%%%%%%%%%%%%%%%%%%%%%%%%%%%%%%%%%%%%%%%%%

The contribution to $\epsilon_m$ from coagulation comes from a double integral over pairs of aerosols $(u_1, u_2)$. Without loss of generality, we assume $u_1 > u_2$, and we only need to consider cases in which $u_2 < v$ but $u_1+u_2 > v$, because coagulation between two aerosols larger than $v$ contributes nothing to $\epsilon_m$. We can then distinguish between 6 different regions in the remaining part of the $(u_1, u_2)$ plane, as is depicted in fig.\ \ref{fig:u1_u2_zones}. The approximate contribution from each of these regions  to $\epsilon_m$ is summarized in table\ \ref{table:zones_u1_u2}. 

For example, we shall consider the case where the volumes of the coagulating of both coagulating particles is smaller but of the same order as $v$. As a consequence, the volume ``produced" from the coagulation is $v$, the area in the $(u_1,u_2)$ plane is of order $v^2$. Since both coagulating particles are of the same order as $v$, their total number will be of order $n(v)^2$, i.e., we assume it to be constant in the integral, which is why the total contribution from this region comes out to be $\sim \beta_0 \times v^2 \times n(v)^2 \times v = \beta_0 v^3 n(v)^2 $. 

\begin{table}[ht]
\centering % used for centering table
\begin{tabular}{c c c c c c} % centered columns (6 columns)
\hline\hline %inserts double horizontal lines
Number & Definition & Area & Number of & Volume  & Total contribution \\ [0.5ex] % inserts table 
              &                 &          &  Particles  &  contributed &  to $\epsilon_m$ \\
%heading
\hline % inserts single horizontal line
1 & $u_{1,2} \lessapprox v$ 				& $v^2$ 	& $n(v)^2$ 		& $v$ & $\beta_0 v^3n(v)^2$ \\ % inserting body of the table
2 & $u_1 \gtrapprox v$, $u_2 \lessapprox v$  	& $v^2$ 	& $n(v)^2$ 		& $v$ & $\beta_0 v^3n(v)^2$ \\
3 & $u_1 \gg v$, $u_2 \lessapprox v$ 			& $\infty$ 	& $n(v)n(u_1)$ 	& $v$ & $\beta_0 v^2n(v)\int_{v}^{\infty} n(u_1) du_1$ \\
4 & $u_1 \lessapprox v-v_1$, $u_2 \gtrapprox v_1$ & $v_1^2$	 & $n(v)n(v_1)$ 	& $v$ & $\beta_0 v_1^2vn(v)n(v_1)$ \\
5 & $u_1 \gtrapprox v$, $u_2 \gtrapprox v_1$ 	& $v_1v$ 	& $n(v)n(v_1)$ 	& $v_1$ & $\beta_0 v_1^2vn(v)n(v_1)$ \\
6 & $u_1 \gg v$, $u_2 \gtrapprox v_1$ 		& $\infty$ 	& $n(v_1)n(u_1)$ 	& $v_1$ & $\beta_0 v_1^2n(v_1)\int_{v}^{\infty} n(u_1) du_1$ \\ [1ex] % [1ex] adds vertical space
\hline %inserts single line
\end{tabular}
\caption{Regions in the $(u_1, u_2)$ plane contributing to $\epsilon_m$. For each of the regions in fig.\ \ref{fig:u1_u2_zones}, we crudely approximate (up to numerical factors) the area of the region in the $(u_1,u_2)$ plane, the number of coagulating particles $n(u_1)n(u_2)$, and the total volume contribution to particles of size $v$ or more. Using these quantities, we estimate the total contribution to $\epsilon_m$ as {\em $\beta_0 \times$ Area $\times$ number of particles $\times$ volume contributed}.  We deal with infinities by later assuming a power law behavior for $n(v)$ and calculate the integral.}
\label{table:zones_u1_u2} % is used to refer this table in the text
\end{table}

We now assume a power-law solution, $n(v)=Av^{-p}$. It is then apparent that one must have $p>1$, otherwise the total number of aerosols $\int n(v)dv$ will diverge. We also assume $p<2$, otherwise the integral $\int n(v)vdv$ converges to a finite volume and not one which can increase linearly with time, as we expect in the steady state.

Under the aforementioned assumptions to total contribution to $\epsilon_m$ from the regions described in table\ \ref{table:zones_u1_u2}) comes out to be 
\begin{equation}
\epsilon_m(v) \approx \beta_0 v^3A^2v^{-2p} + \beta_0 v_1^2vA^2v_1^{-p}v^{-p} \approx \beta_0 A^2v^{3-2p},
\end{equation}
where we have also assumed $v_1 \ll v$ for the second approximation.

If we further assume that $\epsilon_m(v)$ is independent of $v$, we find that $p=3/2$ and $A \approx \sqrt{\epsilon_m / \beta_0}$. As we shall see in the more rigorous treatment described in \S\ref{sec:fullnocondense}, this result is correct up to a numerical factor.

The last result can be further improved by adding the leading correction obtained when including condensation. To do so, we guess a solution in the form $n(v)=Av^{-p}+Cv^{-q}$. This leads to two additional terms in the expression for $\epsilon_m$, which are
\begin{equation}
\epsilon_m(v) \approx \beta_0 A^2v^{3-2p} + 2\beta_0 ACv^{3-p-q}  + \int_{v}^{\infty} Au^{-p} \sigma u^{\gamma} du + A v^{1-p} \sigma v^\gamma ,
\end{equation}
where the second term is a second-order coagulation term, the third term is the first-order condensation term describing the growth of particles larger than $v$, while the last term describes the flux of particles which through condensation become larger than $v$. Note that the last two terms are similar in size if the third term converges which we require anyway. This requirement is equivalent to $\gamma <1/2$. Without this requirement, an upper cutoff for the distribution is necessary to limit the condensation onto the large volume tail.

The ``zeroth" order equation obtained from requiring that $\epsilon_m$  must be independent of $v$ in steady state leads to the same values of $A$ and $p$ found above. The ``first" order equation obtained leads to 
\begin{equation}
0 \approx  \beta_0 Cv^{3-p-q}  + {\sigma}v^{1+\gamma-p}.
\end{equation}
Since this equation holds for all $v$, we find that $q=2-\gamma$ and $C \approx -{\sigma}/{\beta_0}$.  As we shall see in \S\ref{sec:fullSolution}, a more rigorous treatment yields the same correction up to a $\gamma$-dependent numerical factor.

\subsection{Full Solution with no condensation correction}
\label{sec:fullnocondense}

\cite{Klett1975} solved the fixed coalescence cross-section case without while neglecting the condensation equation using Laplace transform, and found that 
\begin{equation}
n\left(v\right)\approx\sqrt{\frac{\epsilon_{m}}{2\pi\beta_{0}}}v^{-3/2} ,
 \end{equation}
 for large values of $v$. Here we show a shorter and more intuitive way to reach the same solution. In the next subsection, we will add the correction term obtained when adding the condensation at large $v$'s (when the condensation term is necessarily small). 

We define  $\epsilon\left(x\right)$ to be the dimensionless rate of change of the total dimensionless volume of particles of dimensionless size $x$ or more (with ``dimensionless" hereafter omitted), that is:
\begin{equation}
\epsilon(x) \equiv \int_{x}^{\infty}\frac{\partial}{\partial\tau}y(\tilde{x})\tilde{x}d\tilde{x} .
\end{equation}

Since we neglect condensation, this increase in the total volume is due to two ``types" of particle coalescence. First, two particles with a volume smaller than $x$ can combine into one particle with a volume larger than $x$. Second, a particle with a volume larger than $x$ can combine with a particle with a volume smaller than $x$. Together, we find that the total volume change due to coalescence is 
\begin{equation}
\epsilon\left(x\right)=\int_{\xi_{1}=0}^{x}\int_{\xi_{2}=x-\xi_{1}}^{x}y\left(\xi_{1}\right)y\left(\xi_{2}\right)\left(\xi_{1}+\xi_{2}\right)d\xi_{1}d\xi_{2}+2\int_{\xi_{1}=0}^{x}\int_{\xi_{2}=x}^{\infty}y\left(\xi_{1}\right)y\left(\xi_{2}\right)\xi_{1}d\xi_{1}d\xi_{2}
 \end{equation}
which simplifies to
\begin{equation}
\label{eq:tempeq}
\epsilon\left(x\right)=2\int_{\xi_{1}=0}^{x}\int_{\xi_{2}=x-\xi_{1}}^{\infty}y\left(\xi_{1}\right)y\left(\xi_{2}\right)\xi_{1}d\xi_{1}d\xi_{2} .
 \end{equation}

Next, we assume a power law solution, i.e., $y=Bx^{-p}$,
and that $1<p<2$. The assumption $p>1$ is necessary to ensure a finite total number of particles, while the assumption $p<2$
  is necessary to ensure an infinite total volume. The latter is necessary because the total volume is a monotonically increasing function of time.
Together with eq.\ \ref{eq:tempeq}, we find after some algebra that
\begin{equation}
\epsilon\left(x\right)=2B^{2}x^{3-2p}\frac{1}{p-1}\int_{\eta=0}^{1}\eta^{1-p}\left(1-\eta\right)^{1-p}d\eta .
 \end{equation}

In steady state, $\epsilon$
  should be independent of x. Therefore $p=3/2$, and
\begin{equation}
\epsilon=2B^{2}\frac{1}{3/2-1}\int_{\eta=0}^{1}\eta^{-1/2}\left(1-\eta\right)^{-1/2}d\eta=4\pi B^{2} .
 \end{equation}
Thus, the steady-state solution to eq.\ \ref{eq:DimensionlessEquation} is
\begin{equation}
y\left(x\right)=\sqrt{\frac{\epsilon}{4\pi}}x^{-3/2} .
 \end{equation}

If we define $\epsilon_{m}$
  to be the rate of change of volume in the original equation, then it can easily be seen that
\begin{equation}
\label{eq:EpsilonAndEpsilonm}
\epsilon_{m}=S_{0}v_{2}\epsilon .
 \end{equation}

Therefore, the solution to  eq.\ \ref{eq:DimensionEquation} with no condensation (for large aerosol volume) is
\begin{equation}
n(v)=\sqrt{\frac{\epsilon_{m}}{2\pi\beta_{0}}}v^{-3/2},
 \end{equation}
which is the same as the solution found by \cite{Klett1975}.

Incidentally, we can use equation \ \ref{eq:EpsilonAndEpsilonm} and the definition of $v_2$ to get
\begin{equation}
\label{eq:SigmaElimination}
\sqrt{S_{0}\beta_{0}}\left(\frac{\epsilon_{m}}{S_{0}\epsilon}\right)^{\left(1-\gamma\right)}=\sigma.
 \end{equation}

This result will be used later in order to eliminate $\sigma$
  from our results, by introducing the dimensionless $\epsilon$ whose value depends on $\gamma$. 
  This is useful because under most physical scenarios, $\sigma$ is unknown, but $\epsilon_m$ is determined from various chemical or physical processes.

\subsection{Full Solution with condensation correction}
\label{sec:fullSolution}

The next step is to generalize the calculation described in \S\ref{sec:fullnocondense} by adding the first-order correction term associated with condensation. With the latter term, the equation describing the increase of the total mass of particles larger than $x$ now includes two additional terms 
\begin{equation}
\label{eq:EpsilonSmallGamma}
\epsilon\left(x\right)=2\int_{\xi_{1}=0}^{x}\int_{\xi_{2}=x-\xi_{1}}^{\infty}y\left(\xi_{1}\right)y\left(\xi_{2}\right)\xi_{1}d\xi_{1}d\xi_{2}+\int_{x}^{\infty}\sqrt{2}\xi^{\gamma}y\left(\xi\right)d\xi+\sqrt{2}x^{\gamma}y\left(x\right) x.
 \end{equation}
The second term in the equation describes volume change due to condensation on particles of volume $x$ or larger. The third term describes particles of volume slightly less than $x$, which in a unit time grow to become larger than $x$ due to condensation.

We will now look for a solution of the type $y=Bx^{-p}+Dx^{-q}$, where $q>p$. We now plug this solution into the equation for $\epsilon$, and neglect small powers of x. For $\gamma < 1/2$, the equation for the highest power of $x$ is unaffected by the new term $Dx^{-q}$, so we still have $p=3/2$
  and $B=\sqrt{\epsilon/(4\pi)}$
  as before.

The equation we get for the next highest power of $x$ is (after some algebra):
\begin{equation}
0=2Dx^{3-p-q}\left[\frac{1}{q-1}+\frac{1}{p-1}\right]\int_{0}^{1}\eta^{1-p}\left(1-\eta\right)^{1-q}d\eta+\sqrt{2}x^{1+\gamma-p}\left(1+\frac{1}{p-1-\gamma}\right) .
 \end{equation}

\begin{figure}[h]
\centering
\includegraphics[width=0.6\columnwidth]{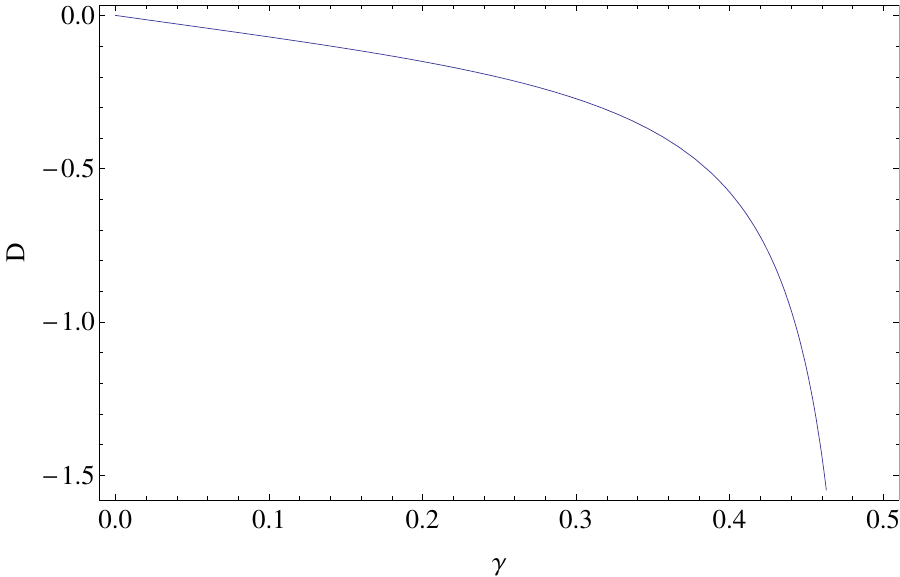}
\caption{$D$ vs. $\gamma$ for the case $1/2 > \gamma > 0$.}
 \label{fig:Dgraph}
\end{figure}

As before, this equation should hold for all values of $x$ (in the limit $x\gg1$). Equating the power law indices of $x$ gives us:
\begin{equation}
\label{eq:q_small_gamma}
2-\gamma=q.
 \end{equation}
 This explains why $\gamma$ should be smaller than $1/2$ for this solution to be valid. 
Next, we can equate the coefficients. Once we plug in $p = 3/2$ and $q = 2-\gamma$, we find
\begin{equation}
\label{eq:Dequation}
2D\left[\frac{1}{1-\gamma}+\frac{1}{1/2}\right]\int_{0}^{1}\eta^{-1/2}\left(1-\eta\right)^{\gamma-1}d\eta=-\sqrt{2}\left(1+\frac{1}{1/2-\gamma}\right).
 \end{equation}
This equation can be solved for $D$. For example, in the important case where $\gamma={1}/{3}$ we have $\int_{0}^{1}\eta^{-1/2}\left(1-\eta\right)^{\frac{1}{3}-1}d\eta=\sqrt{\pi}\Gamma\left(1/3\right) / \Gamma\left(5/6\right) \approx4.20655 $
such that eq.\ \ref{eq:Dequation} gives $D(\gamma = 1/3) = -0.3362$. For other values of $\gamma$, see fig.\ \ref{fig:Dgraph}.

%%%%%%%%%%%%%%%%%%%%%%%%%%%%%%%%%%%%%%%%%%%%%%%%%%%%%%%%%%%%%%%%
%%%%%%%%%%%%%%%%%%%%%%%%%%%%%%%%%%%%%%%%%%%%%%%%%%%%%%%%%%%%%%%%

To summarize, the solution of eq.\ \ref{eq:DimensionlessEquation}  is of the form (for $0 \le \gamma < 1/2$):
\begin{equation}
y\left(x\right)=\sqrt{\frac{\epsilon}{4\pi}}x^{-3/2}+Dx^{-\left(2-\gamma\right)} .
 \end{equation}
For the specific case $\gamma=\frac{1}{3}$, we obtain
\begin{equation}
y\left(x\right)=\sqrt{\frac{\epsilon}{4\pi}}x^{-3/2}-0.336x^{-5/3}.
 \end{equation} 
Using the physical quantities, this specific solution becomes
\begin{equation}
n\left(v\right)=\sqrt{\frac{\epsilon_{m}}{2\pi\beta_{0}}}v^{-3/2}-0.475\frac{\sigma}{\beta_{0}}v^{-5/3}.
 \end{equation}
In many cases, we are given $\epsilon_m$, i.e., the sulfuric acid formation rate, instead of $\sigma$, its equilibrium number density. 
We can therefore use eq.\ \ref{eq:SigmaElimination} and find
\begin{equation}
\label{eq:SolutionGammaThird}
n\left(v\right)=\sqrt{\frac{\epsilon_{m}}{2\pi\beta_{0}}}v^{-3/2}-0.475\beta_{0}^{-1/2}S_{0}^{-1/6}\left(\frac{\epsilon_{m}}{\epsilon}\right)^{2/3} v^{-5/3}.
 \end{equation}

A numerical solution of the equation (described in the appendix) shows that $\epsilon(\gamma = 1/3) \approx 3.296$. This is perhaps the most important result in the present work since it describes the leading two terms in the distribution of aerosols in the atmosphere. The analytic form of the second term is described here for the first time. 

In principle, it is possible to add higher order corrections as a power series of $x$, where each term's power decreases by $1/6$.

\subsection{Full Solution with condensation correction for $\gamma > 1/2$}
\label{sec:gammaBiggerHalf}

The solution described in \S \ref{sec:fullSolution} is valid only for $\gamma < 1/2$, because it was assumed that $q > p$ such that $2-\gamma > 3/2$  (using eq.\ \ref{eq:q_small_gamma}). Therefore, the larger $\gamma$ case should be solved separately. As we shall see below, this has a major effect on the solution---the leading power of $x$ changes.

First, because $\epsilon(x)$ diverges, instead of working with $\epsilon(x)$ we will work with $\epsilon(x,x_3)$ which is the rate of change in the total mass between $x$ and $x_3$. In steady state, it clearly vanishes, even though the integral above $x_3$ diverges. 

%Thus, we write $\epsilon$:
%\begin{equation}
%\label{eq:EpsilonLargeGamma}
%\epsilon\left(x\right)=2\int_{\xi_{1}=0}^{x}\int_{\xi_{2}=x-\xi_{1}}^{\infty}y\left(\xi_{1}\right)y\left(\xi_{2}\right)\xi_{1}d\xi_{1}d\xi_{2}+\sqrt{2}x^{\gamma}y\left(x\right) x+\int_{x}^{x_{3}}\sqrt{2}\xi^{\gamma}y\left(\xi\right)d\xi .
%\end{equation}
%It is the same as eq.\ \ref{eq:EpsilonSmallGamma}, except for the upper bound $x_{3}$
% which  was introduced in the second integral. This is required to avoid the divergence of $\epsilon$. As will be evident below, the value of $x_3$ will not be important for sufficiently large $x_3$.

Again we assume a power-law solution of the type $y=Bx^{-p}$, and write the solution in the form
\begin{equation}
\label{eq:Fdef}
\epsilon\left(x,x_3\right)=F(x)-F(x_{3}),
\end{equation}
where the expression for $F(x)$ is exactly the same as the expression we had for $\epsilon(x)$ in eq.\ \ref{eq:EpsilonSmallGamma}, where we drop the upper bound wherever it diverges. Using this form, we find after integration that
\begin{equation}
F(x)=2B^{2}x^{3-2p}\frac{1}{p-1}\int_{\eta=0}^{1}\eta^{1-p}\left(1-\eta\right)^{1-p}d\eta+\sqrt{2}Bx^{1+\gamma-p}-\sqrt{2}B\frac{1}{1+\gamma-p}x^{1+\gamma-p}.
\end{equation}

Since $\epsilon(x,x_3)$ should vanish, we must require $F(x)$ to be independent of $x$.
However, since $p \ne 3/2$, the only way for $F(x)$ to be independent of $x$ is for the $x$-dependent terms to cancel each other out. This implies that the exponents are equal, such that, $2-\gamma=p$.
Last, we can find the pre-factor $B$ by requiring the coefficient of $x^{3-2p}$ (or $x^{1+\gamma-p}$) to vanish as well. The result is
\begin{equation}
\label{eq:SolutionLargeGammaDimensionless1}
y\left(x\right) = B  x^{2-\gamma} = \left[ \sqrt{2}\frac{\left(1-\gamma\right)^{2}}{2\gamma-1}\frac{1}{\int_{\eta=0}^{1}\eta^{\gamma-1}\left(1-\eta\right)^{\gamma-1}d\eta} \right] x^{2-\gamma} .
\end{equation}

For instance, the solution to eq.\ \ref{eq:DimensionlessEquation} for $\gamma=2/3$ and large values of $x$ is approximately given by $y=0.230 x^{-4/3}$. In terms of $v$, the solution to eq.\ \ref{eq:DimensionEquation} for the above case is
\begin{equation}
n\left(v\right)=0.325\frac{\sigma}{\beta_{0}} v^{-4/3}.
\end{equation}

\begin{figure}[h]
\centering
\includegraphics[width=0.6\columnwidth]{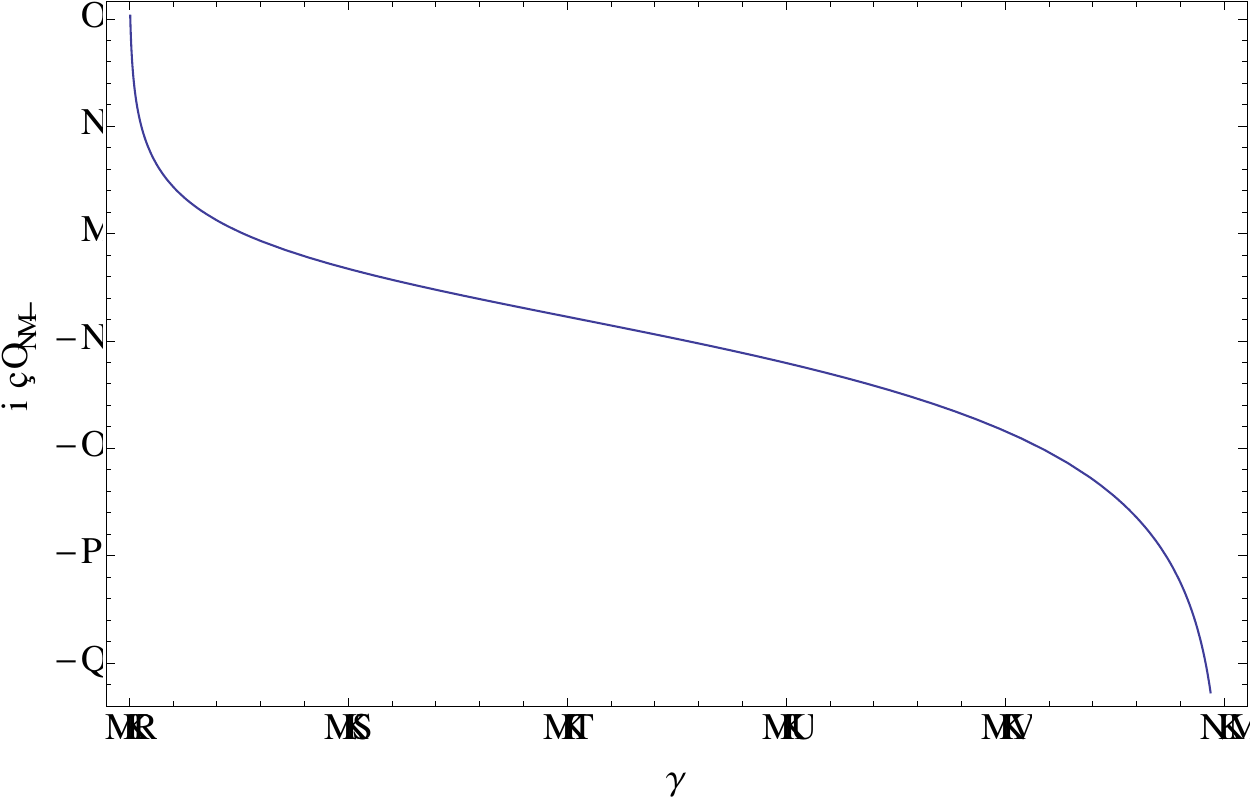}
\caption{$B$ vs. $\gamma$ for the case $1 > \gamma > 1/2$.}
 \label{fig:Bgraph_large_gamma}
\end{figure}

For other values of $\gamma$ see fig.\ \ref{fig:Bgraph_large_gamma}. Unlike the previous case, of $\gamma < 1/2$, the condensation term's contribution to $\epsilon_m$ formally diverges. This implies that without additional physics introducing a large volume cutoff, such as dry or wet deposition, or a dependence of $\gamma$ on the aerosol volume, the solution is no longer physical. It also implies that we cannot eliminate $\sigma$ from the above results. 
 
\subsection{Full Solution with 2$^\mathbf{nd}$ condensation correction for $\gamma > 1/2$}

As for the small $\gamma$ case described in \S\ref{sec:fullSolution}, it is possible to derive a higher order correction to the lowest order solution described in \S\ref{sec:gammaBiggerHalf} above.  To do so, we can repeat \S\ref{sec:gammaBiggerHalf} under the assumption that the solution is of the more general form $y=Bx^{-p}+Dx^{-q}$ (where $q>p$), and calculate $F(x)$ defined in eq.\ \ref{eq:Fdef}.

\begin{figure}[h]
\centering
\includegraphics[width=0.6\columnwidth]{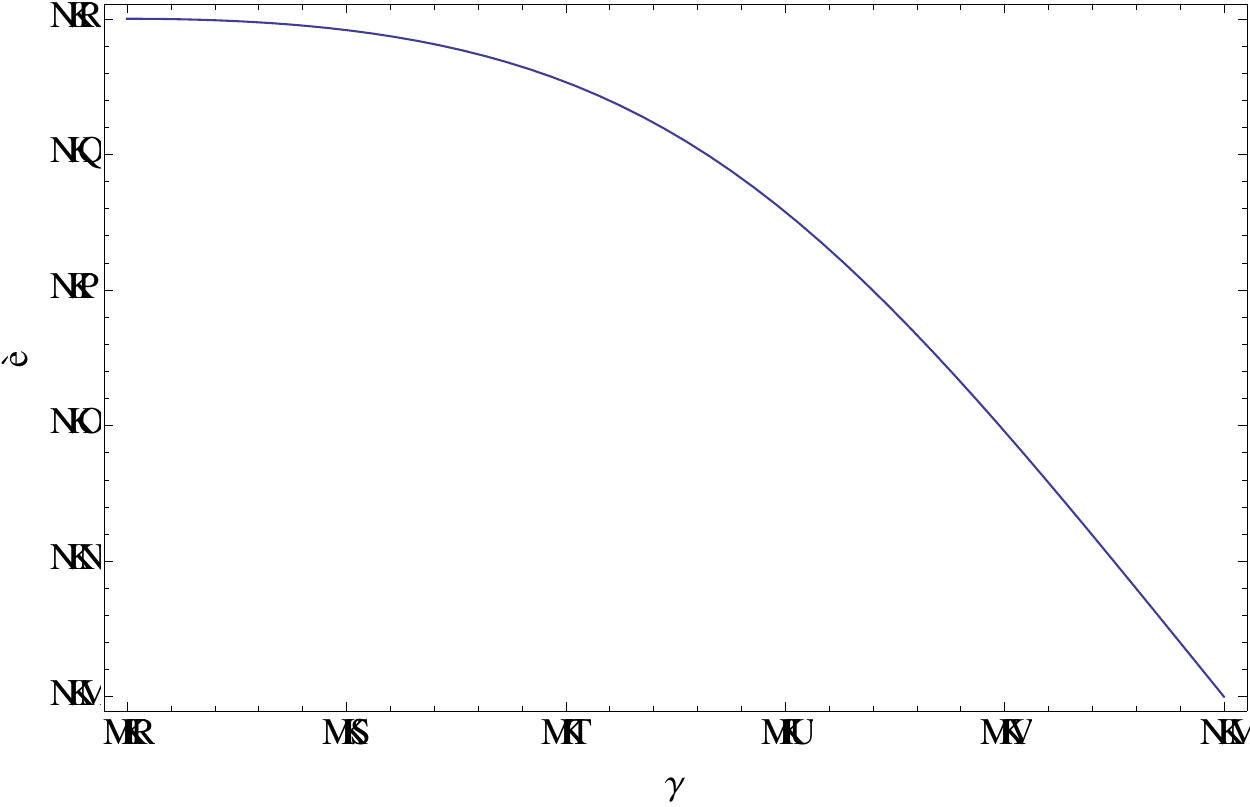}
\caption{$q$ vs. $\gamma$ for the case $1 > \gamma > 1/2$.}
 \label{fig:qvsgamma}
\end{figure}

We then require $F(x)=0$ and compare the leading terms in powers of $x$. The lowest order term was described above (see eq.\ \ref{eq:SolutionLargeGammaDimensionless1}). The next order gives,  
\begin{eqnarray}
0 = 2BDx^{3-p-q}\int_{0}^{1}\eta^{1-p}\frac{1}{q-1}\left(1-\eta\right)^{1-q}d\eta+ \nonumber \\
+2BDx^{3-p-q}\int_{0}^{1}\eta^{1-q}\frac{1}{p-1}\left(1-\eta\right)^{1-p}d\eta+ \nonumber \\
+\sqrt{2}Dx^{1+\gamma-q}-\sqrt{2}D\frac{1}{1+\gamma-q}x^{1+\gamma-q}
\end{eqnarray}
The requirement on the exponents gives $3-p-q=1+\gamma-q$, however, it follows straightforwardly from $p = 2 - \gamma$ found in the first order solution. Consequently, we only obtain new information by requiring the sum of the coefficients to vanish. Since $D$ cancels out, the constraint becomes
\begin{equation}
\sqrt{2}\frac{1}{1+\gamma-q}-\sqrt{2}=2B\left[\frac{1}{q-1}+\frac{1}{p-1}\right]\int_{0}^{1}\eta^{1-p}\left(1-\eta\right)^{1-q}d\eta.
\end{equation}
This equation can be solved numerically for $q$, given the previously calculated values of $p$ and $B$. For example, the solution for the case $\gamma=2/3$ is $q\approx1.47$. Solutions for other values of $\gamma$ are described in fig.\ \ref{fig:qvsgamma}.

%%%%%%%%%%%%%%%%%%%%%%%%%%%%%%%%%%%%%%%%%%%%%%%%%%%%%%%%%%%%%%%%
%%%%%%%%%%%%%%%%%%%%%%%%%%%%%%%%%%%%%%%%%%%%%%%%%%%%%%%%%%%%%%%%

%%%%%%%%%%%%%%%%%%%%%%%%%%%%%%%%%%%%%%%%%%%%%%%%%%%%%%%%%%%%%%%%
%%%%%%%%%%%%%%%%%%%%%%%%%%%%%%%%%%%%%%%%%%%%%%%%%%%%%%%%%%%%%%%%

%%%%%%%%%%%%%%%%%%%%%%%%%%%%%%%%%%%%%%%%%%%%%%%%%%%%%%%%%%%%%%%%
\begin{figure}[h]
\centering
\includegraphics[width=0.6\columnwidth]{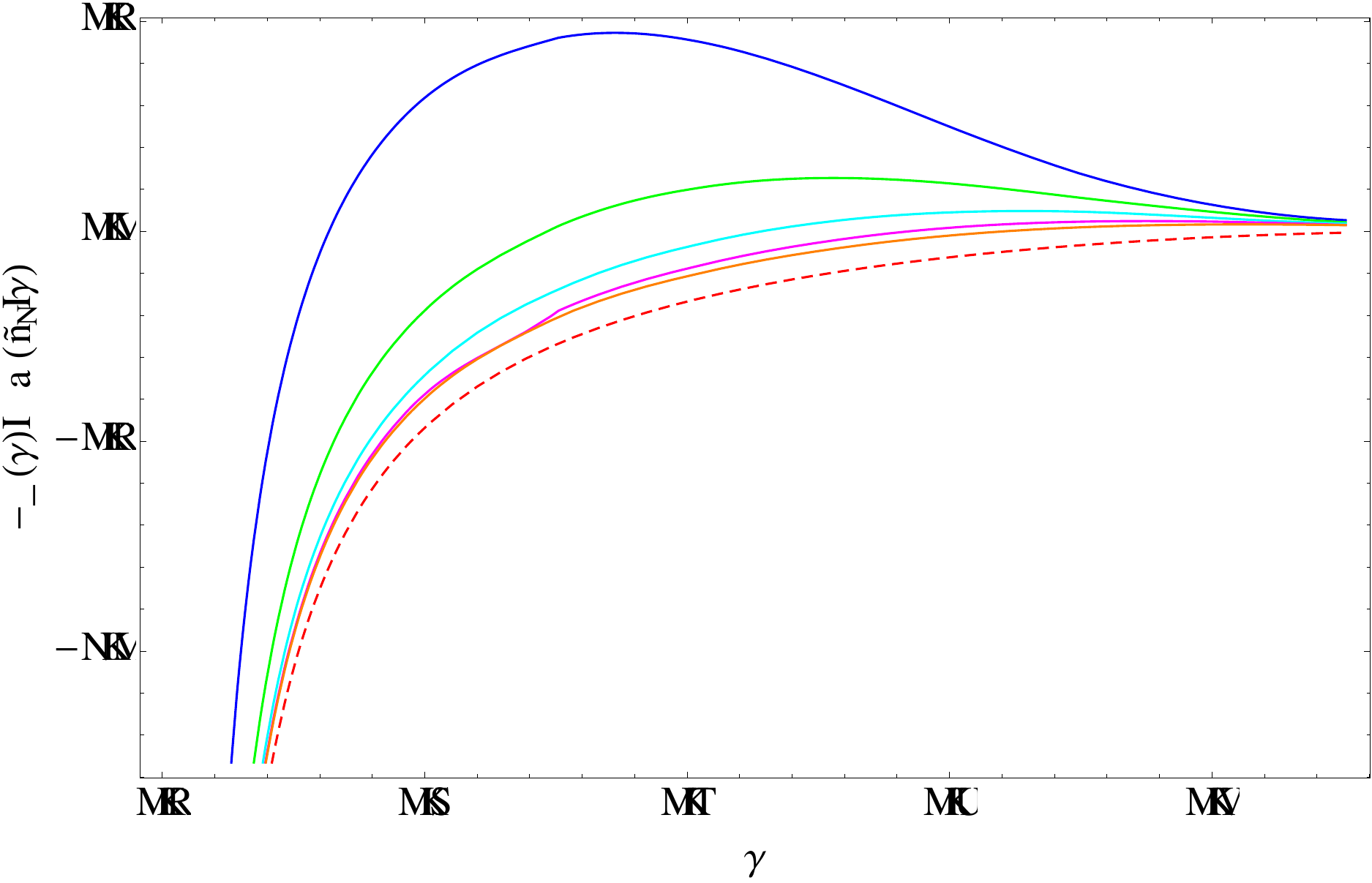}
\caption{(color onloine) $D$ vs. $\gamma$ and $-B$ vs. $\gamma$ for the case $1 > \gamma > 1/2$. The dashed line depicts $-B$ while the solid lines are from top to bottom: $x_1=10,1,0.1,0.01,0.001$.}
 \label{fig:Dgraph_large_gamma}
\end{figure}
%%%%%%%%%%%%%%%%%%%%%%%%%%%%%%%%%%%%%%%%%%%%%%%%%%%%%%%%%%%%%%%%

The value of $D$ is found numerically, and it is found to have a rather strong dependence on $x_{1}$. Note that $D$ can be negative. The numerical  simulations also seem to indicate that for $x_1 \rightarrow 0$, $D(x_1) \rightarrow -B$. This can be seen in fig.\ \ref{fig:Dgraph_large_gamma}.

In terms of the equation\ \ref{eq:DimensionEquation}, we have the following solution for the case $\gamma=2/3$:
\begin{equation}
n\left(v\right)=0.325\frac{\sigma}{\beta_{0}} v^{-4/3}+D(x_1 = v_1/v_2)\sqrt{\frac{2S_{0}}{\beta_{0}}}\left(\frac{\sigma^{2}}{S_{0}\beta_{0}}\right)^{0.706}v^{-1.471} .
\end{equation}

\section{The condensation limit}
\label{sec:CondensationLimit}

Until now we studied the limits where coagulation is much more important than condensation. We now concentrate on the opposite limit, where condensation is much more important. This describes, for example, the growth of small aerosols. We begin by describing the pure condensation limit, and then continue by crudely adding the coagulation as a correction. 

\subsection{The condensation solution without coagulation corrections}
\label{sec:nocoagulation}

After discarding both coagulation terms in eq.\ \ref{eq:DimensionlessEquation}, the steady state equation can be solved analytically. Eq.\ \ref{eq:DimensionlessEquation} simply becomes
\begin{equation}
\sqrt{2}\frac{d}{dx}\left[x^{\gamma}y\left(x\right)\right]=\delta(x-x_{1}).
 \end{equation}

The general solution is $y=Cx^{-\gamma}$, where $C$ can be found by the ``boundary condition" at $x=x_{1}$, which yields
\begin{equation}
y=Cx^{-\gamma}=\frac{1}{\sqrt{2}}x^{-\gamma}.
 \end{equation}

\subsection{The condensation solution with a coagulation correction}

The pure condensation solution can be extended by adding the first order correction arising from coagulation. This introduces two terms appearing in eq.\ \ref{eq:DimensionlessEquation} which are estimated by using the ``zeroth order" solution. The first coagulation term is roughly $\int_{0}^{x}y\left(\tilde{x}\right)y\left(x-\tilde{x}\right)d\tilde{x} \approx y(x)y(x)x \approx x^{1-2\gamma}$. The second coagulation term is roughly $-2y(x) \approx x^{-\gamma}$. Therefore, assuming $\gamma < 1$, the first coagulation term will be much smaller than the second coagulation term for all $x\ll 1$, and we shall not take it into account in the calculation that follows. The steady state dimensionless equation can still be solved analytically. The equation is
\begin{equation}
\sqrt{2}\frac{d}{dx}\left[x^{\gamma}y\left(x\right)\right]=-2y\left(x\right)+\delta(x-x_{1}).
\end{equation}
Its general solution is given by
\begin{equation}
y=Cx^{-\gamma}\exp\left[-\sqrt{2}\frac{1}{1-\gamma}x^{1-\gamma}\right].
 \end{equation}
Using the ``boundary condition" at $x=x_{1}$, and assuming $x_{1}\ll1$,  we obtain:
\begin{equation}
y\left(x\right) \approx \frac{1}{\sqrt{2}}x^{-\gamma}\exp\left[-\sqrt{2}\frac{1}{1-\gamma}x^{1-\gamma}\right],
 \end{equation}
or in terms of eq.\ \ref{eq:DimensionEquation}
\begin{equation}
\label{eq:SolutionCondensation}
n\left(v\right) \approx \frac{S_{0}}{\sigma}v^{-\gamma}\exp\left[-\sqrt{2S_{0}\beta_{0}}\frac{1}{\sigma}\frac{1}{1-\gamma}v^{1-\gamma}\right].
\end{equation}
In the case $\gamma<1/2$, we can eliminate $\sigma$ from this result, and give it instead in terms of $\epsilon_m$, using  eq.\ \ref{eq:SigmaElimination}. This gives
\begin{equation}
n\left(v\right)\approx\sqrt{S_{0}/\beta_{0}}\left(\frac{\epsilon_{m}}{S_{0}\epsilon}\right)^{\gamma-1}v^{-\gamma}\exp\left[-\sqrt{2}\left(\frac{\epsilon_{m}}{S_{0}\epsilon}\right)^{\gamma-1}\frac{1}{1-\gamma}v^{1-\gamma}\right],
\end{equation}
where $\epsilon$ is $\gamma$-dependent, and must be calculated numerically.

Note that this solution can easily be generalized to the case where a particle-loss term $-\lambda n\left(v\right)$ is added to eq.\ \ref{eq:DimensionEquation}. 

\section{Comparison with the numerical solutions}
\label{sec:Numerical}

Although there is no exact analytic solution for the full coagulation-condensation equation, a full solution can be obtained numerically, as is described in the Appendix.

\begin{figure}[h]
\centering
\includegraphics[width=0.8\columnwidth]{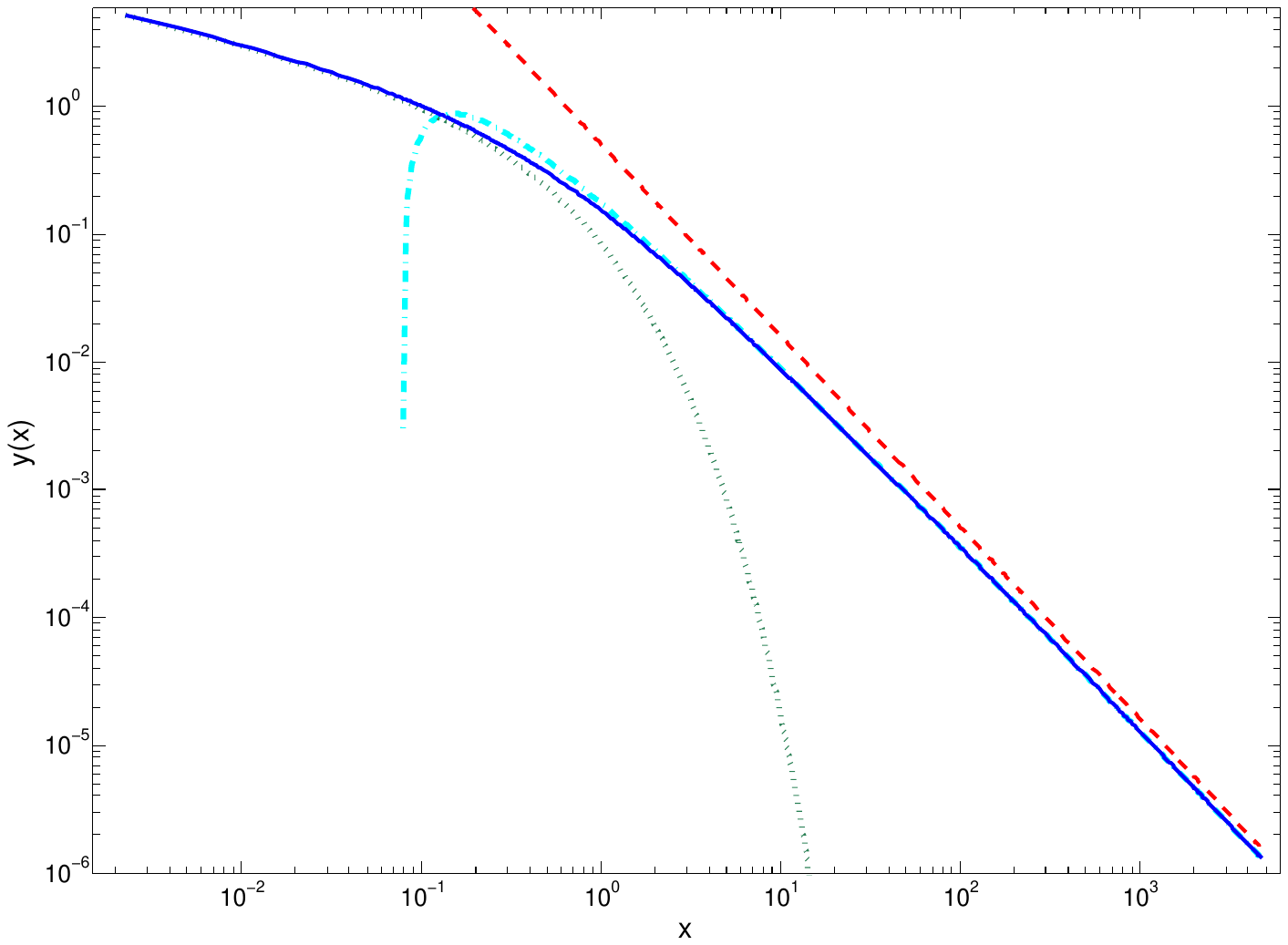}
\caption{(color online) Comparison between the numerical solution (blue) and the analytic approximations derived here, for the case $\gamma = 1/3$. The zeroth order solution describing coagulation and its first order correction from condensation are depicted with the red and cyan lines respectively, which are valid for large $x$'s. The zeroth order condensation solution with its first order coagulation correction valid for small $x$'s is described by the green line. }
 \label{fig:analytic_and_numerical_solutions_gamma_third}
\end{figure}

The full solution can then be used to check the quality of the analytical approximations obtained above. 
Fig.\ \ref{fig:analytic_and_numerical_solutions_gamma_third} plots the full numerical solution and the analytical approximations obtained for the case $\gamma=1/3$. As is evident, the two analytic approximations for the $x\gg1$ and the $x\ll 1$ limits are relatively accurate. In fact, at $x \sim 0.5$ both solutions are only 30\% off the exact solution. 

%%%%%%%%%%%%%%%%%%%%%%%%%%%%%%%%%%%%%%%%%%%%%%%%%%%%%%%%%%%%%%%%
%%%%%%%%%%%%%%%%%%%%%%%%%%%%%%%%%%%%%%%%%%%%%%%%%%%%%%%%%%%%%%%%

\section{Discussion}
\label{sec:Discussion}

One of the interesting implications of the solution to the coagulation/condensation equation is that it helps us understand how the aerosol distribution will change under different perturbations to the CN nucleation rate and to the amount of available condensable gas. 

It is known from numerical simulations that perturbing the nucleation rate $S_0$ will have an appreciable effect on the density of small CNs, whose growth is governed by condensation. However, the effect is going to be small for large CCNs once coagulation becomes important \citep{svensmark2013}. 

On the other hand, a perturbation to the amount of available condensable material $\epsilon_m$, which is usually sulfuric acid, will affect both the number density of small CNs and of the larger CCNs. 

With the result of eq.\ \ref{eq:SolutionGammaThird} and eq.\ \ref{eq:SolutionCondensation}, this behavior can be understood and quantified analytically. 

It is probably possible to obtain better theoretical solutions in the coagulation limit by taking the next term in the power series. However, this will probably not introduce any new physical consequences to the solution---unlike the first order correction to the solution (see eq.\ \ref{eq:SolutionGammaThird}) which added a dependence on the nucleation rate.

We also solved the case $\gamma > 1/2$, which corresponds to high condensation rates for large particles (see eq.\ \ref{eq:SolutionLargeGammaDimensionless1}). The zeroth order solution was again improved by finding the first correction term. Since the rate of change in the total volume diverges, the solution has no physical meaning without introducing, for example, a large size cutoff. 

As mentioned in  \S\ref{sec:generalkernel}, another interesting aspect is that the solution can be straightforwardly generalized to describe kernels of the type $\beta\left(v,\tilde{v}\right)=\beta_{1}v^{\alpha}\tilde{v}^{\alpha}$ through the definition $\gamma^{\prime}=\gamma-\alpha$. This allows us to describe more realistic aerosol growth in the real atmosphere. 

This implies that in the condensation limit the generalized solution is 
\begin{equation}
y\left(x\right) \approx \frac{1}{\sqrt{2}}x^{-\gamma-2\alpha}\exp\left[-\sqrt{2}\frac{1}{1-\gamma-\alpha}x^{1-\gamma-\alpha}\right],
\end{equation} while for the coagulation limit we find 
\begin{equation}
y(x) = B x^{-p-\alpha} + Dx^{-q-\alpha},
\end{equation}
where $B, D, p$ and $q$ are defined with $\gamma'$ instead of $\gamma$.

For example,
In the limit were coagulation is through hydrodynamic capture of particles in the Stokes regime \citep{Klett1975}, we can expect $\alpha  = 2/3$.

%\note{emphasize the cases $\gamma=1/3$ and $\gamma=2/3$, but remember that $\alpha$ shifts over the value of $\gamma$, therefore other $\gamma$'s are also important.}

\section*{Acknowledgments}

This research project was supported by the I-CORE Program of the Planning and Budgeting Committee
and The Israel Science Foundation (Center No. 1829/12). NJS also thanks the IBM Einstein Fellowship support by the IAS.

  \def\aj{Astron.\ J.}
\def\apj{Ap.\ J.}
\def\apjs{Ap.\ J.\ Supp.}
\def\apjl{Ap.\ J.\ Lett.}
\def\mnras{Mon.\ Not.\ Roy.\ Astro.\ Soc.}
\def\aap{Astron.\ Astrophys.}
\def\araa{Ann.\ Rev.\ Astron.\ Astrophys.}
\def\pasj{Pub.\ Astron.\ Soc.\ Japan}
\def\apss{Astrophys.\ Sp.\ Sci.}
\def\nar{N.\ Astron. Rev.}
\def\nat{Nature}

\bibliography{AnalyticAerosols_for_arxiv}

\section*{Appendix - Numerical Solution}
\label{sec:Appendix}

Here we describe the numerical solution of eq.\ \ref{eq:DimensionEquation}. Since the equation is the time-independent limit of the full equation (eq.\ \ref{eq:FullCCE}) one method of solving the time independent equation is to solve the time dependent one and let the system relax to its steady state solution. However, we choose a second approach which reduces the CPU usage considerably. 

First, we discretize the volume of the particles, as $v_{1},v_{2},v_{3},\dots$.
We arbitrarily choose $v_{i}=i$ for all $i$, and denote the number of particles of size $i$ by $n_{i}$. 
Doing so, we now observe that the time-dependent equation for $n_{1}$ is:
\begin{equation}
\frac{\partial n_{1}}{\partial t}+\sigma(v_{1})^{1/3}n_{1}=0-\beta_{0}n_{1}\chi_{0}+S_{0}.
 \end{equation}
 Note that the 0 term denotes the fact that the particles with the smallest volume cannot be formed through the coagulation of smaller particles. 
For $n_{k+1}$, the equation is:
\begin{equation}
\frac{\partial n_{k+1}}{\partial t}-\sigma(k)^{1/3}n_{k}+\sigma(k+1)^{1/3}n_{k+1}=\frac{1}{2}\beta_{0}\sum_{i=1}^{k}n_{i}n_{k+1-i}-\beta_{0}n_{k+1}\chi_{0}.
 \end{equation}
 
 Since the coagulation coefficient is assumed to be constant, the dynamics of $n_{k+1}$ depends only on $n_1\dots n_k$ and on $\chi_0$ which was found to be $\sqrt{{2S_{0}}/{\beta_{0}}}$ in \S\ref{sec:dimensionless}. As a consequence, the equations can be solved sequentially from $n_1$. 
 
Note also that the solution can be accelerated by using FFT to compute the convolution term. In this fashion we easily reach $n_{2,000,000}$ in our simulations using just a single CPU.

\end{document}